\documentclass[12pt]{article}
 \input{epsf}
 \usepackage{epsfig}
\usepackage{amsfonts}

\begin{document}

 \title{A time-dependent brane in a cosmological background}

\author{  Nicolaos Toumbas \\
{\em Department of Physics } \\
{\em University of Cyprus  } \\
{\em Nicosia, CY-1678, Cyprus } \\
and \\ 
Jan Troost
\\ {\em  Laboratoire de Physique Th\'eorique } \\
{\em Ecole Normale Sup\'erieure\thanks{Unit{\'e} mixte  du
CNRS et de l'Ecole Normale Sup{\'e}rieure,
UMR 8549.} } \\  {\em 24, Rue Lhomond } \\
{\em Paris 75005,  France }
\thanks{troost@lpt.ens.fr; Preprint LPTENS-04/37;
Research 
partially supported by the EEC under the contracts HPRN-CT-2000-00122,
HPRM-CT-2000-00131. }
   }

\maketitle

 \abstract{We study a moving D-brane in a time-dependent background. There is
   particle production both because of non-trivial cosmological evolution, and
 by closed string emission from the brane that gradually decelerates due to
a gain in mass. The particular model under study is a $D0$-brane in an
   $SL(2,R)/U(1)$
cosmology -- the techniques used extend to other backgrounds.}

\section{Introduction}
The study of time-dependent phenomena in string theory has gained momentum,
partly under the influence of impressive improvements in the measurement of
cosmological data (e.g. the detailed mapping of the cosmic microwave 
background). The euclidean framework that underlies perturbative string
theory does not lend itself easily to interpretation in time-dependent
settings. It would seem that any progress we can make in continuing 
hard euclidean results in string theory into the Lorentzian domain, and
in interpreting them sensibly, is worthwhile reporting.
(See e.g. 
\cite{Gutperle:2003xf}\cite{Strominger:2003fn}\cite{Schomerus:2003vv}\cite{Fredenhagen:2003ut}\cite{Kutasov:2004dj}\cite{Nakayama:2004yx}.)

In this letter, we choose to study an exact cosmological solution to string
theory (see e.g. \cite{Kounnas:1992wc}\cite{Cornalba:2002nv}\cite{Craps:2002ii}),
exhibiting non-trivial time-dependence in the closed string sector, and
we supplement it with a non-trivial moving brane that solves the equations
of motion of classical open string field theory. Exact results in the purely
closed string background can be obtained from the study of a particular coset
conformal field theory, $SL(2,R)/U(1)$, while exact results in the open string
sector can be obtained from analytically continuing boundary states of the
euclidean model 
(see \cite{Ribault:2003ss}\cite{Israel:2004jt}\cite{Fotopoulos:2004ut}). Two
of the challenges we 
face is to identify the correct
analytic continuation, as well as to interpret correctly the time-dependent
physics.

In section \ref{cosm} we discuss the exact cosmological background, and
in part \ref{bulkprod}
the particle production due to the time-evolution. In section \ref{addsource}
we set up the formalism that will allow us to add the effects due to 
emission of on-shell closed strings from an extra source. Then, in 
section \ref{brane}, we discuss the particular D-brane source we add, and
the closed string emission it generates. We wind up with conclusions
in section \ref{conclusions}.

\section{Cosmological Background}
\label{cosm}
We study the cosmology with metric and dilaton
\begin{eqnarray}
ds^2 &=& |k \alpha'| \frac{dudv}{1-uv} \nonumber \\
e^{2 \Phi} &=& \frac{e^{2 \Phi_0}}{1-uv} 
\end{eqnarray}
which is known to correspond to an exact two-dimensional conformal field
 theory,
the gauged Wess-Zumino-Witten model $SL(2,R)_k/U(1)$, where 
%the generator
%that is gauged is hyperbolic, and where 
we consider negative level $k$.
As such, it can be combined with flat space factors, and the euclidean 
two-dimensional black hole coset $SL(2,R)/U(1)$ to create a critical
 (super)string theory background \cite{Kounnas:1992wc}. 

In the following, we will make good use
of a set of alternative coordinate systems, which can be defined as
follows (we will neglect the overall factor in the metric on occasion):
\begin{eqnarray}
u &=& X^1-X^0 \quad ; \quad v = X^1+X^0 \nonumber \\
ds^2 &=& \frac{-(d{X^0})^2+(d{X^1})^2}{1-({X^1})^2+({X^0})^2}
\end{eqnarray}
which is useful to study global properties of the spacetime and exhibits the conformally flat nature of the cosmology (up to a non-trivial dilaton factor);
\begin{eqnarray}
u &=& e^x \sinh t \quad ; \quad v= - e^{-x} \sinh t \nonumber \\
ds^2 &=& -dt^2 + \tanh^2 t dx^2 \nonumber \\
e^{2 \Phi} &=& \frac{e^{2 \Phi_0}}{\cosh^2 t} ,
\end{eqnarray}
which is a useful coordinate system for comparison to the euclidean 2-dimensional black hole
and
\begin{eqnarray}
u &=& e^x \tilde{t} \nonumber \\
v &=& -e^{-x} \tilde{t} \nonumber \\
ds^2 &=& \frac{1}{1+ \tilde{t}^2} (-d \tilde{t}^2+ \tilde{t}^2 dx^2)
\nonumber \\
|\tilde{t}| &=& e^{\eta} \nonumber \\
ds^2 &=& \frac{e^{2 \eta}}{1+ e^{2 \eta}} (- d \eta^2 + dx^2),
\end{eqnarray}
which makes manifest the relation of the cosmology to flat space in Milne-like
coordinates near the origin $\tilde{t}=0$.
\label{2dcosmologypenrose}
\begin{figure} 
\begin{center}
 \epsfxsize=5cm
\epsfbox{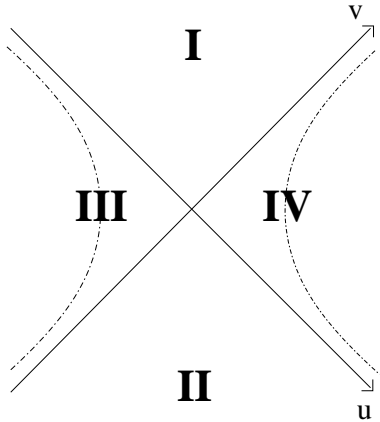}
\caption{\em The Penrose diagram for the two-dimensional cosmology \em }
\end{center}
\end{figure}

The spacetime Penrose diagram is drawn in figure (1). 
 There are four regions separated by horizons at $u=0$ and $v=0$. Regions I
 and
 II are free of any singularities; they can be thought of as expanding and 
collapsing cosmologies respectively. Region I is defined by $uv < 0, \ X^0>0$,
 while region II by $uv<0, \ X^0<0$. The coordinates defined by equation (3)
 (and also (4)) are suitable to parameterize these regions.
In terms of the $(x,t)$ coordinates, the horizon corresponds to the surface
 $t=0$, and for the cosmological region I $0 \le t < \infty$. 
In regions III and IV ($uv>0$) (naively)
there are timelike naked singularities. No observer in region I can ever meet
 the singularities; they are simply outside the future lightcone of such an
 observer. However, it is clear that the singularities can influence the
 cosmological evolution in region I. In this sense region I is similar 
to a big bang cosmology. The crucial difference is that near $t=0$ the
 spacetime is smooth having no curvature singularity and initial data 
can be defined on this surface. In the $(x,t)$ coordinates, the metric
 near $t=0$ is given approximately by the flat metric in Milne-like coordinates and the
 string coupling is constant, while as $t \to \infty$ the metric approaches
 the flat Minkowski metric and the string coupling constant vanishes 
exponentially. Thus, asymptotically the cosmological region I becomes
 identical
 to a timelike linear dilaton background.

We will also make use of the boost symmetry of the cosmology: $(u,v)\to (e^{-x_0}u, e^{x_0}v)$. It is easy to see that in the $(x,t)$ coordinates this symmetry corresponds to translations in the spatial $x$-direction. Because of the symmetry, the $x$-momentum is conserved, but due to the time-dependence of the metric there is no conservation law for the energy. 

In the following, our idea will be to concentrate on the future region
of the cosmology, namely region I. This region naturally Wick rotates
 to the euclidean black hole (after
changing also the sign of the level), which is a well-studied euclidean
conformal field theory. We thus concentrate only on this part of the cosmology\footnote{The other
regions, namely (parts of the) regions III and IV, analytically
continue to compact parafermions,
$SU(2)/U(1)$. It would be
interesting to investigate how to interpret these regions and their 
parafermionic analytic continuation (and in particular the resulting
boundary conditions, spectrum, etc). We thank Costas Kounnas for instructive
discussions on these issues.}. 
                 It should be noted that different attitudes can
be taken and have been taken toward the proper definition of the conformal
field theory corresponding to this and other Lorentzian backgrounds. One 
consists in defining the coset strictly in terms of the (hyperbolic) quotient of the
original
group manifold (see 
e.g. \cite{Elitzur:2002rt}\cite{Craps:2002ii}), and a second
consists in following closely techniques of quantum field theory analysis in curved space
for a particular region of the Lorentzian space under study. We follow the
second method since the euclidean conformal field theory
that
results from this analysis is under sufficient control to allow for exact
results. It is also desirable to
study further the direct Lorentzian WZW model (and its quotient), and to 
further compare results obtained through these two complementary methods.

\section{Cosmological Particle Production}
\label{bulkprod}
In this section, we wish to study particle creation in the bulk of the cosmology for a scalar closed string field $\phi$ with quadratic effective Lagrangian
\begin{eqnarray}
{\cal L} &=& \frac{1}{2} \sqrt{-g} e^{- 2 \Phi} (-g^{\mu \nu} 
\partial_{\mu} \phi \partial_\nu \phi - m^2 \phi^2).
\end{eqnarray}
In the region of interest, namely the cosmological region I, the dilaton field can be chosen to be arbitrarily small and so as a first step we ignore terms in the effective action describing string interactions. The analysis in this section is equivalent to a mini-superspace analysis of strings propagating in the cosmology, and the results obtained here should match the exact string theoretic results in the limit of infinite negative level. As we shall argue later, some exact string theory results at finite level can be obtained by suitably analytically continuing the corresponding euclidean black hole CFT results.
The resulting equation of motion is then given by
\begin{eqnarray}
\left(- \frac{1}{\sqrt{-g} e^{-2 \Phi}} \partial_{\mu} (e^{- 2 \Phi} \sqrt{-g}
  g^{\mu \nu} \partial_\nu) +m^2 \right) \phi &=& 0.
\end{eqnarray}

% \subsection*{Norm}

As usual the current density (corresponding to a
complexified
field $\phi$)
\begin{eqnarray}
j_{\mu} &=& e^{- 2 \Phi} \sqrt{-g} (\phi \partial_\mu \phi^{\ast}-
                                    \phi^{\ast} \partial_\mu \phi)
\end{eqnarray}
is conserved 
and can be used to define an invariant (modified Klein-Gordon) inner product
\begin{eqnarray}
(\phi_1,\phi_2) &=& -i \int_{\Sigma} \sqrt{-g} e^{-2 \Phi}
(\phi_1 \partial_{\mu} \phi_2^{\ast}-\partial_\mu \phi_1
\phi_2^{\ast}) d \Sigma^{\mu}
\end{eqnarray}
(in notation borrowed from \cite{BD}). Due to conservation of the
current,
the inner product is independent of the choice of the spacelike slice $\Sigma$
we use to compute it. The modes of the scalar field can thus be normalized using this inner product, and the resulting norm is independent of time.
%\subsection*{Solutions EOM}

We now turn to solving the equation of motion. In terms of the $(x,t)$ coordinates, the wave-operator is given by 
\begin{eqnarray}
\Box &=&  \frac{1}{|k| \alpha'}
(\partial_t^2 + (\coth t + \tanh t) \partial_t - \coth^2 t \partial_x^2),
\end{eqnarray}
where we reintroduced the level $k$ of the conformal field theory.
It is useful to define the parameter $y=\sinh^2 t$. In terms of $y$, the wave-operator becomes 
\begin{eqnarray}
\Box &=& \frac{1}{|k| \alpha'} (4 \partial_y y(1+y) \partial_y- \frac{1+y}{y} \partial_x^2).
\end{eqnarray}
The equation of motion can then be written as the following eigenvalue equation
\begin{eqnarray}
(\Box + m^2) \phi=0.
\end{eqnarray}
Because $f(y)=e^{-ipx} y^{i|p|/2} \phi(x,y)$ satisfies a
hypergeometric equation with parameters
\begin{equation}
a=j+1-i|p|/2,\ b=-j-i|p|/2, \ c=-i|p|+1; \ \ z=-y, 
\end{equation}
we can easily find the solutions and expand the field in terms of them
\begin{eqnarray}
\phi &=& \int_{-\infty}^{+\infty} \frac{dp}{2 \pi} d e^{ipx}
y^{-i|p|/2} F(j+1-i|p|/2,-j-i|p|/2,-i|p|+1;-y) 
\nonumber \\
& & + d^{\ast} e^{-ipx} y^{i|p|/2} 
F(j+1+i|p|/2,-j+i|p|/2,i|p|+1;-y). \label{decom}
\end{eqnarray}
Here
\begin{eqnarray}
m^2 &=& -\frac{ p^2}{|k| \alpha'} -4 j(j+1)/(|k|\alpha') \nonumber \\
    &=& -\frac{ p^2}{|k| \alpha'}+\frac{1+4 s^2}{|k| \alpha'} \label{onshell} 
%\nonumber \\
%    &=& -\frac{8}{\alpha'} L^0_{SL(2,R)_k/U(1)},
\end{eqnarray}
can be interpreted as a mass squared.
We assume 
that we are dealing with continuous representations of
$SL(2,R)$, or in other words, with delta-function normalizable wave-functions,
for which $j=-\frac{1}{2}+is$ where $s \ge 0$.

Having solved the equation of motion, we turn to defining quantum fields at
early and  late times and obtain the corresponding vacua. This requires to identify positive and negative frequency modes
at early and late times.
\subsection*{Early modes}
In the early flat region (in Milne-like coordinates)
near $t=0$ (and $t$ positive), the modes defined by the decomposition in equation (\ref{decom}) are approximately given by
\begin{eqnarray}
\phi_p &=& d
e^{ipx} t^{-i|p|} 
\nonumber \\ 
& & + d^{\ast} e^{-ipx}
t^{i|p|}.
\end{eqnarray}
This leads us to identify the normalized, positive frequency, early modes 
\begin{eqnarray}
u^{e}_p &=& \frac{1}{\sqrt{2|p|}} e^{ipx} 
y^{-i|p|/2} F(j+1-i|p|/2,-j-i|p|/2,-i|p|+1;-y) 
\nonumber 
\end{eqnarray}
satisfying $(u_p^e,u_{p'}^e) = 2 \pi \delta(p-p')$.
 (The index 'e' refers to early modes, while the index 'l' will refer
to late modes.)
We quantize the field in terms of these early modes by defining annihilation and
creation operators as follows
\begin{eqnarray}
\phi^{e} &=& \int \frac{dp}{2 \pi} \left( a_p^e u_p^{e}
+ {a_p^e}^{\dagger} {u^{e}_p}^{\ast} \right).
\end{eqnarray}    
The operators satisfy the usual commutation relations
\begin{eqnarray}
{[} a_p^e , {a_{p'}^e}^{\dagger} {]} &=& 2 \pi \delta(p-p').
\end{eqnarray}
The early vacuum $|e\rangle$ is annihilated by all $a_p^e$, $a_p^e|e\rangle=0$.

A few comments are in order.
\begin{itemize}
\item 
As we already remarked, the metric near $t=0$ is given approximately by the flat metric in Milne-like coordinates. The early modes are positive and negative frequency with respect to the conformal time $\eta$, defined by $|t|=e^{\eta}$ near $t=0$. Thus the early vacuum is approximately given by the conformal Milne vacuum (see \cite{BD}) in the region near $t=0$.
\item
The corresponding frequencies near $t=0$ are entirely determined by the momentum independently of the mass of the field. This fact can be explained as follows. The modes correspond to particles whose coordinate energy (conjugate to the time $t$)
scales like $E_e^2 - \frac{p^2}{t^2} = m^2$ at early times.
Thus as $t\to 0$, for fixed mass and momentum, the energy is dominated by the momentum dependent piece,
$E_e \sim |p|/t$, which explains the result.
\item
Upon the analytic continuation $|p|\to i|n|$, and a change in the sign of the
level, the positive frequency mode continues to the normalizable mode of the
euclidean black hole (see e.g. \cite{Ribault:2003ss}). 
Thus correlators in the early vacuum are expected to be related to the
corresponding 
euclidean CFT correlators upon a suitable continuation.
\end{itemize} 
\subsection*{Late modes}
We define the late-vacuum at time $t=+\infty$ as follows.
To obtain positive and negative frequency modes at late times, 
it is more convenient to
write the solutions in the form
\begin{eqnarray}
\phi &=& \int \frac{dp}{2 \pi}
 b e^{ipx} y^{-j-1} F(j+1+i|p|/2,j+1-i|p|/2,2j+2;-y^{-1})
\nonumber \\
& & + b^{\ast} e^{-ipx} y^j F(-j-i|p|/2,-j+i|p|/2,-2j;-y^{-1}).
\end{eqnarray}
As $t \to + \infty$, the wavefunctions are given by 
\begin{eqnarray}
\phi &=& \int \frac{dp}{2 \pi} 
b e^{ipx} (4)^{1/2+is} e^{- t} e^{- i2st}
+ b^{\ast} e^{-ipx} (4)^{1/2-is} e^{- t} e^{+ i 2st},
\end{eqnarray}
which leads to the identification of the normalized, positive frequency, late times mode:
\begin{eqnarray}
u_p^{l} &=&  \frac{1}{\sqrt{4s}}  e^{ipx}
 y^{-j-1} F(j+1+i|p|/2,j+1-i|p|/2,2j+2;-y^{-1}).
\end{eqnarray}
This choice of modes defines the late vacuum following the steps described above. 
We make the following observations:
\begin{itemize}
\item
In the (x,t) coordinates and at late times, the metric asymptotes to the flat 
Minkowski metric but the dilaton is linear in time. The string coupling
constant is exponentially suppressed in time. The behavior of the late times
modes as $t \to \infty$ matches onto the mode solutions for a timelike linear
dilaton background \cite{Craps:2002ii}.
\item
From the behavior of the modes as $t \to \infty$, we see that the parameter $s$ defined by equation (\ref{onshell}) determines the frequency of the modes or the energy of the corresponding particles at late times, $\omega=2s$.
\end{itemize}

\subsection*{Particle creation}
The early and late times modes can be related via a
transformation in the arguments of the corresponding hypergeometric functions
(notations and conventions are summarized in the appendix):
\begin{eqnarray}
u_{p}^{l} &=& \frac{1}{\sqrt{4 s}} 
e^{ipx} y^{-j-1} F(j+1+i|p|/2,j+1-i|p|/2,2j+2;-y^{-1})
\nonumber \\
&=&  \frac{1}{\sqrt{4  s}} 
e^{ipx}{[} y^{i |p|/2} \Gamma(\frac{-i|p|,2j+2}{1+j-i|p|/2,j+1-i|p|/2})
\nonumber \\
& &  
F(j+1+i|p|/2,-j+i|p|/2,+i|p|+1;-y) \nonumber \\
& & - y^{- i |p|/2}
\Gamma(\frac{i|p|+1,-i|p|,2j+2}{1-i|p|,1+j+i|p|/2,1+j+i|p|/2})
\nonumber \\
& & 
F(j+1-i|p|/2,-j-i|p|/2,-i|p|+1;-y){]} 
\end{eqnarray}
leading to the identification of the Bogoljubov coefficients
\begin{eqnarray}
{\alpha_{p}}^{p'} &=& \delta(p-p') \alpha(p) 
\nonumber \\
{\beta_{p}}^{p'} &=& \delta(p+p')  \beta(p)
\end{eqnarray}
with
\begin{eqnarray}
{\alpha(p)} &=& - \sqrt{\frac{ |p|}{2 s}}  
\Gamma(\frac{i|p|+1,-i|p|,2j+2}{1-i|p|,1+j+i|p|/2,1+j+i|p|/2})
\nonumber \\
{\beta(p)} &=&   \sqrt{\frac{ |p|}{2 s}}
 \Gamma(\frac{-i|p|,2j+2}{1+j-i|p|/2,j+1-i|p|/2}).
\end{eqnarray}
The Bogoljubov transformation can be written as follows
\begin{eqnarray}
u^l(p) &=&  \alpha(p)u^e(p)+\beta(p){u^e(-p)}^\ast 
\nonumber \\
u^e(p) &=&  \alpha(p)^{\ast}u^l(p)-\beta(p){u^l(-p)}^\ast 
\end{eqnarray}
leading to the following relation in terms of the annihilation and creation operators
\begin{equation}
a_p^l=\alpha(p)^\ast a_p^e - {\beta(p)}^{\ast} ({{a_{-p}^e}})^\dagger.
\end{equation}
As usual the unitarity relation $|\alpha|^2-|\beta|^2 = 1$
is satisfied. Since the early vacuum is not annihilated by $a^l_p$, it can be thought of as an excited state of the late vacuum, and therefore an observer at infinity concludes that particles have been created.

The total number of particles (for a given single field) produced from evolving the vacuum at early times
to late times is given by
\begin{equation}
\langle e| N^l |e\rangle = 2\pi\delta(0)\int {dp \over 2\pi} |\beta(p)|^2.
\end{equation}
The infinite factor $2 \pi \delta(0)$ 
can be thought of as a volume factor and arises because of translational invariance in the spatial $x$-direction. Therefore, $\int dp |\beta(p)|^2$ determines the number of particles produced per unit volume. 
In our case, we obtain
\begin{eqnarray}
| \beta |^2 
%&=& 
%\frac{8 |p|}{s}
%| \Gamma(\frac{-i|p|,1+2is}{-i|p|/2+1/2+is,1/2+is-i|p|/2})|^2 
%\\
%&=& \frac{8 |p|}{s} \frac{4 s^2 \cosh \pi(|p|/2-s)^2}{|p| 2s \sinh \pi 2s
%| \sinh \pi |p|}
%\nonumber \\
&=& \frac{ \cosh^2 [\pi(|p|/2-s)] }{ \sinh (2\pi s) \sinh (\pi|p|)}.
\end{eqnarray}
Two limits of phase space parameters are interesting: 
\begin{itemize}
\item
In the limit of large momentum (and fixed mass), 
using the on-shell condition equation (\ref{onshell}), the energy is given approximately by $\omega =2s \sim |p|$ and so for large momentum $|\beta(p)|^2 \sim e^{-2\pi |p|}=e^{-2\pi \omega}$ leading to an exponential suppression in the production of high momentum particle pairs. 
\item
In the small momentum limit, the energy is given approximately by the mass
$\omega=2s \sim m_{eff}=(|k|\alpha'm^2/4-1)^{1/2}$ and so $|\beta(p)|^2 \sim 
\frac{ \coth{(\pi m_{eff}/2)}}{2 \pi |p|}$.
 This leads to a divergence in the integral arising from the infrared region. For a massive field, the divergence is logarithmic. The surprising result is that the infrared divergence persists independently of the mass of the field. 
\end{itemize}

Thus the particle spectrum produced is dominated by low momentum quanta. 
%The infrared divergence indicates that the early perturbative vacuum is unstable. As we shall discuss later, the problem persists in string theory. 
%In fact in perturbative string theory, backreaction becomes more severe due to the Hagedorn density of states in the theory.   
We can understand the result better by computing the rate of pair creation for a given mode.
S-matrix elements between the early times vacuum and late times particle states are characterized by the quantity $\gamma$ (or its complex conjugate) defined by \cite{BD}
\begin{equation}
\gamma^\ast(\omega)=-{\beta(p) \over \alpha(p)^\ast}={\Gamma(2j+1)\Gamma^2(-j-i|p|/2) \over \Gamma(-2j-1) \Gamma^2(1+j-i|p|/2)}. \label{gamma}
\end{equation}
Thus the absolute value of $\gamma(\omega)$ characterizes the rate of creation of a given particle 
pair \footnote{Conservation of $x$-momentum requires that pairs of particles of equal and opposite momentum are produced.}. This is given by 
\begin{equation}
|\gamma(\omega)|^2 = {\cosh^2 [\pi(|p|/2-s)] \over \cosh^2 [\pi(|p|/2+s)]}. \label{rate}
\end{equation}
For large momentum, $\omega=2s \sim |p|$ and the rate is suppressed $|\gamma|^2 \sim e^{-2\pi \omega}$. For small momentum though, the rate is of order one: for $p \sim 0$, $|\gamma|^2 \sim 1$. Thus small momentum particles are readily produced.

It can be easily seen that when we analytically continue $|p| \to i|n|$, the
amplitude in equation (\ref{gamma}) becomes identical to the mini-superspace result for
the reflection amplitude ${\cal{R}}(j,n)$ of a closed string mode (with
quantum numbers $(j,n)$) in the euclidean black hole geometry
\cite{Ribault:2003ss}. The mini-superspace result is equivalent to the $k \to
\infty$ limit of the relevant euclidean CFT two-point function. Thus at finite
level $k$, we expect the exact answer for the rate amplitude to be obtainable
from the euclidean CFT reflection amplitude upon the same analytic
continuation accompanied also by a change in the sign of the level. Similar
results have been conjectured to hold in the cases of time-like bulk and
boundary Liouville theories \cite{Gutperle:2003xf}\cite{Strominger:2003fn}. 
Assuming that this is true, the exact amplitude is given by equation (28) multiplied
by a pure 
phase factor given by \cite{Ribault:2003ss}
\begin{equation}
{\nu_b}^{i\omega}{\Gamma (1 + ib^2\omega)\over \Gamma ( 1-ib^2\omega)}
\end{equation}
with $\nu_b=\Gamma(1-b^2)/\Gamma(1+b^2)$ and 
$b^2= -1/(|k|+2)$. This implies that the rate (\ref{rate}) determined by the
modulus of $\gamma$ is exact 
even at finite level $k$ (except for an overall rescaling of
the metric due to the level shift). 
Thus the analysis of the particle spectrum we obtained can be carried through
beyond the mini-superspace
 analysis at finite (shifted) level $k$.

The results we found above indicate clearly that cosmological particle
 production leads to significant backreaction with a breakdown of perturbation
 theory. From the two dimensional field theoretic point of view, the
 problematic region is the infrared small momentum region. The divergence we
 found in the produced particle density can be associated to an instability of
 the early cosmological vacuum. For a massive scalar field, the divergence is
 milder 
 and since 
low momentum particles are produced at a rate of order one in string
 units \footnote{The time-scale for the breakdown of perturbation theory is
 set by the 
string scale. No matter how small the initial value of the string coupling is,
 once
 enough particles are produced, they will lead to significant backreaction.}, the
 space quickly fills with such non-relativistic particles. In other words, the
 space rapidly fills with dust-like matter, and presumably
 it eventually collapses. 
From the string theoretic point of view, the
 situation is more
 complicated and the resulting backreaction more severe. In this case, we
 need to
 take into account the dependence of the effective mass on the momenta along
 the extra
 spatial directions and the Hagedorn density of string states at large
 energy
 (and oscillator level). We discuss this case next.

\subsection*{Embedding in string theory}
We saw that for a fixed (cosmological) mass of a given closed string field, and at large energy, the rate of particle 
production is either exponentially suppressed as a function of the energy
$|\gamma|^2 \sim e^{-2\pi\omega}$
 or of order one $|\gamma|^2 \sim 1$, depending on whether the momentum along the $x$-direction is large or small.
When we further add
appropriate
factors to build a critical string theory, e.g. by adding a euclidean black
hole conformal field theory (at level $|k|+4$) and a twenty-two dimensional flat space factor, 
the rate of particle production at large energy will be enhanced due to the
exponentially 
large Hagedorn density of closed string states: 
$\rho(\omega)= e^{ \omega  \over \sqrt{|k|} T_H}$. 
In this example, $T_H={1 \over 2\pi}\sqrt{6 \over D-2}={1 \over 4\pi}$, and so
the total 
rate of particle production $\rho(\omega)|\gamma|^2$ at large energy is 
exponentially divergent\footnote{We do not expect the details of the spins and 
polarizations of the closed string fields along the extra directions to alter
this conclusion.},
for low-momentum modes. For high-momentum modes, the exponential suppresion of
the amplitude is sufficient to render the particle production rate finite, for
sufficiently large $|k|$, i.e. for sufficiently weakly curved
cosmologies.\footnote{We would like to thank Costas Kounnas for correcting an
erroneous statement in the first version of our paper.}

%In fact it is easy to identify
%regions in phase space parameters where the particle production rate at large
%energy is exponentially divergent, irrespectively of the details of the string
%theory model we construct. One such
% region is the region where 
%the oscillator level $N$, and hence the energy, of closed string states that
%are produced are very large while all momenta are
%kept fixed (and small). 
%
We will generically run into this phenomenon in cosmological models where
 we can identify a regime in parameter space in
which the rate of particle creation (per field) is of order one (e.g. for low
 momentum, high masses),
while the degeneracy of states (as a function of the square root of the oscillator level)
grows exponentially with an exponent which is (twice) the
inverse of the closed string Hagedorn temperature
(see e.g. \cite{BZ} for the factor of two). The
type of matter produced in this regime consists of heavy, almost 
non-relativistic strings and is similar to the ``tachyon matter'' produced
during the decay
 of unstable branes \cite{Sen:2002in}\cite{Lambert:2003zr}\cite{Sen:2003xs}. Within a time-scale of order the
string length, enough particles are produced to lead to large backreaction and
a breakdown
 of perturbation theory. Understanding the fate of the cosmology then requires
 understanding
 strings at very high densities. 

The breakdown of perturbation theory indicates that non-perturbative effects
are 
important in understanding the fate of these cosmological models. One
expectation is that such non-perturbative effects will cut-off the
exponentially growing density of string states. Given this, particle
production 
in string cosmology can be turned into an advantage in constructing realistic
models of
 inflationary cosmology since such particle production may result in
 re-heating the 
universe \cite{Gubser:2003vk}\cite{Friess:2004zk} after an early inflationary phase. 

It is useful (and sometimes appropriate) to think of the inverse of the
exponent 
in the rate amplitude as a cosmological temperature. 
In the example we studied, the cosmological temperature is momentum dependent,
and
 for small momenta, it is clearly far above the Hagedorn temperature
since then the cosmological temperature is infinite. Another
recent example where a similar phenomenon was observed is the $0<k<2$ cosmology of \cite{Hikida:2004mp}.
In time-like bulk Liouville theory, the situation is slightly better in that
the
 cosmological temperature
is only double the Hagedorn temperature (still leading to large
backreaction)\cite{Strominger:2003fn}.
% This is the regime in our case as well for high $x$-momentum.

String theory seems to naturally produce cosmological solutions
in which the cosmological temperature is larger than the Hagedorn temperature,
leading to large backreaction\footnote{We thank Dan Israel for discussions on this topic.}. 
One can ask whether there are ways to lower the
cosmological temperature, or to heighten the Hagedorn temperature. We
note that the cosmological temperature depends on the choice of initial state
and initial surface. The leading (mini-superspace) behavior may also receive higher curvature
corrections which at strong curvature could significantly alter 
the cosmological production rate\footnote{However, as we indicated above, by using the connection between
the amplitude for pair creation and the exact euclidean reflection amplitude, one can
show
that no such correction occurs for the particular
 cosmology that we studied -- except for the effects due to an overall
rescaling of the metric.}. 
Heightening the Hagedorn temperature can
be achieved in non-critical strings \cite{Karczmarek:2003xm} (although the
linear 
dilaton behavior would
need to be short-cut by a Liouville-like potential to avoid strong coupling
problems). It would be interesting to understand better the generic constraints
on string theory that lead to these
 UV
 catastrophes in cosmology.

Note that our vacuum
% also
 generates a lot of particles in the IR regime (at
small $x$-momentum $p$). This indicates that we would need to modify the definition
of the early cosmological state in the infrared, for a proper time evolution.
 Indeed, as we argued the onset of the instability occurs at an early timescale.
It would be interesting to understand the physical consequences of such an IR
regularization at early times.

Finally, we add a word on methodology. Implicitly, we have taken the attitude
that a second quantized string theory in this cosmological background can be defined,
such that we can study a collection of second quantized fields with standard methods
developed for quantum fields in curved space. This should be contrasted with computations
in the first quantized approach, in which it would only seem to make sense to ask questions
about the stringy S-matrix. It would be interesting to investigate the formulation of string
field theory on time-dependent backgrounds, and 
to see how these two attitudes should be reconciled.

After these remarks on aspects of bulk string cosmology, we turn to
the study of cosmological effects due to non-perturbative time-dependent
D-branes.   

\section{Particle production in the presence of a source}
\label{addsource}
In this section,
we develop a formalism to take into account simultaneous 
particle creation due to cosmological evolution on the one
hand and due to the presence of a time dependent
source (in a linearized approximation) on the other hand. 
To derive the relevant particle
production amplitude, we follow e.g. \cite{PS}.

The most general solution to the wave equation in the presence of the source can be written as follows
\begin{eqnarray}
\phi(x,t) &=& \phi_{e} (x,t) +
i \int dt' dx'  \sqrt{-g} e^{- 2 \Phi} G_R(x,t;x',t')  \rho(x',t') 
%\nonumber \\
%& & 
% \phi_{early} (x) + i \int d^{d} x \int \frac{d^{d-1} p}{(2 \pi)^{d-1}}
%\frac{1}{2 |E|} \theta(x^0-{x'}^0) .(e^{- i p.(x-x')}-e^{ip.(x-x')})
%\rho(x') \nonumber \\
%& & \int \frac{d^{d-1} p}{(2 \pi)^{d-1}} 
%\frac{1}{\sqrt{2E}}((a+\frac{i}{\sqrt{2E}}
%\tilde{ \rho} (p)) e^{-ip.x} + h.c.) 
\end{eqnarray}
where $\phi_{e}(x,t)$ solves the homogeneous, sourceless wave equation and $G_R(x,t;x't')$ is the retarded propagator satisfying
\begin{eqnarray}
(\Box + m^2) G_R(x,t;x',t') &=& -i \frac{\delta(x-x')\delta(t-t')}{\sqrt{-g} e^{- 2 \Phi}}. 
\end{eqnarray}

The retarded propagator is given by
\begin{eqnarray}
G_R(x,t;x',t') &=& \theta(t-t') \langle e | {[} \phi(x,t) , \phi(x',t'){]} | e \rangle.
\end{eqnarray}
Here $|e \rangle$ refers to the early vacuum. 
Therefore, it can be formally computed in terms of the Wightman functions 
$G(x,t;x',t')=\langle e| \phi(x,t)\phi(x',t') |e \rangle$
and similarly for $G(x',t';x,t)$. In terms of the early modes the Wightman function is given explicitly by
\begin{equation}
G(x,t;x',t')=\int {dp \over 2\pi}{1 \over 2|p|} {e^{ip(x-x')}} {f^j_{|p|}}(t){[}{f^j_{|p|}}(t'){]}^*,
\end{equation}
where we defined ${f^j_{|p|}}(t)= y^{-i|p|/2} F(j+1-i|p|/2,-j-i|p|/2,-i|p|+1;-y)$ with $y=\sinh^2t$.   

Let us suppose that the source starts and stops. Then if we wait until all of the source is in the past, the field solution can be written as follows
\begin{eqnarray}
\phi(x) &=&  \int \frac{d p}{2 \pi}
 \left[a_{p}^{e}+ \frac{i}{\sqrt{2|p|}}
\tilde{\rho}(j(p),p) \right ] u^{e}_p +h.c,
\end{eqnarray}
where $\tilde{\rho}(j,p)$ is the Fourier transform of the source defined 
by\footnote{We summarize completeness/orthogonality relations in
  appendix \ref{formal}.}
\begin{equation}
\tilde{\rho}(j,p) =  \int dt dx \sqrt{-g}e^{-2\Phi} \rho(t,x) 
e^{-i px}   (f^j_{|p|} (t))^{\ast}
\end{equation}
and evaluated at energies and momenta satisfying the on shell condition, eq. (\ref{onshell}). Only the on-shell part
of the source contributes to the late 
quantum field. 
In the case when a localized source persists at very late times, the above formula applies only in the limit $t \rightarrow \infty$, away from the source.

Using the Bogoljubov 
coefficients, we can write the early modes in terms of the late field modes.
%\begin{eqnarray}
%a^e_p &=& \int dp' {\alpha_p}^{p'} a^l_{p'} + {{\beta_p}^{p'}}^{\ast} 
%{a^l_{p'}}^{\dagger}
%\nonumber \\
%a^l_p  &=& \int dp' {{\alpha_p}^{p'}}^{\ast} a^e_{p'} - 
%{{\beta_{p}}^{p'}}^{\ast} {a^{e}_{p'}}^{\dagger}
%\end{eqnarray}
%and
Since
$u^e_p = {\alpha(p)}^{\ast} u^l_{p}- {\beta}(-p) ({u^l_{-p}})^{\ast}$,
we obtain after substitution 
\begin{eqnarray}
\phi(x,t) &=&  \int \frac{d p}{2 \pi} 
 \left((a^e_p+ \frac{i}{\sqrt{2|p|}}
\tilde{\rho}(p) )  ( {\alpha(p)}^{\ast} u^l_{p}- {\beta}(-p) {u^l_{-p}}^{\ast}  )  +h.c. \right ).
\end{eqnarray}
We can now identify the coefficient of $u^l_{p}$ as the late time
annihilation operator including the action of the source to obtain
\begin{eqnarray}
a^{l,\rho}_{p} &=&  
(a^e_p+ \frac{i}{\sqrt{2|p|}} \tilde{\rho}(p)) 
{\alpha(p)}^{\ast} -({a^{e}_{-p}}^{\dagger}-\frac{i}{\sqrt{2|p|}}
\tilde{\rho}(-p)^{\ast}) {{\beta}(-p)}^{\ast}  . 
\end{eqnarray}
We see that we recuperate standard results for cosmological particle
production when the source 
is zero \cite{BD}, and when the early and late
modes in the cosmology are identical, we recuperate standard production of particles
by a classical source \cite{PS}. Finally, we compute the total number of particles produced: 
\begin{eqnarray}
\langle e | N^{l} | e \rangle 
&=& \int {dp \over 2\pi} \left({\cal{V}}| \beta (p)|^2 + 
\frac{1}{2 |p|} |\tilde{\rho}(p) {{\alpha^{\ast}}} (p) +
\tilde{\rho}^{\ast} (-p) {\beta(-p)}^{\ast} |^2 \right).
\label{partprod}
\end{eqnarray}

\section{The brane source}
\label{brane}
We wish to study D0-branes in the cosmology in the mini-superspace approximation. The effective metric for such a brane is given by
\begin{equation}
{ds^2 \over g_s^2}= dudv,
\end{equation}
and so it is flat. Because the effective metric is flat, a D0-brane follows a straight line trajectory in the $(u,v)$ plane. One example is the brane at fixed $X^1=(u + v)/2 =c$. All other brane trajectories can be obtained by performing a boost on this brane  
\begin{equation}
e^{-x_0}u+e^{x_0}v=2c.
\end{equation}
From the point of view of the string frame metric, the branes follow accelerated (non-geodesic) trajectories. The reason is that they experience a force from the changing dilaton background.
In terms of the $(x, t)$ coordinates, the 2-parameter family of the brane trajectories can be written as follows
\begin{eqnarray}
\sinh (x-x_0) \sinh t &=& c.
\end{eqnarray}
Since such probe branes follow straight lines in the $(u,v)$ plane, they seem 
to be oblivious to (naive) space-time singularities\footnote{This behavior is
  directly related to the fact 
that D0-branes in the
two-dimensional black hole pass through singularities without much ado
\cite{Yogendran:2004dm}.}.

The description of such a brane from the point of view of 
the cosmological observer using the $(x,t)$ coordinates is as follows.
For $c > 0$, the brane enters the cosmological region crossing the horizon
$u=0$ at
 finite null time $v_0=2ce^{-x_0}$. In terms of the $(x, t)$ coordinates, this
 event
 occurs at $t=0, \ x\to \infty$. The initial $x$-velocity of the brane is
 infinite\footnote{The proper velocity of the brane is finite as can be seen from the
  behavior of the metric near $t=0$.} (and negative). Subsequently the brane
decelerates until it localizes at $x=x_0$ as $t \to \infty$.
 A similar description can be obtained when $c < 0$. The fact that the brane
 decelerates
 from the point of view of this observer has a natural explanation. The
 dilaton background
 field is time dependent such that it decreases as $t$ goes from zero to
 infinity. During 
late times, the dilaton field decreases exponentially and the metric is
flat -- in the $(x,t)$ 
coordinates the background asymptotically approaches the linear dilaton
background as $t \to \infty$.
 As a result, the brane gets heavier and heavier during cosmological
 evolution. Conservation of
 momentum then implies that it must decelerate.    

Such a brane leads to a delta-function source for closed string fields 
localized on its trajectory. The source is time dependent and part of it
will be on-shell. So we naturally expect on-shell emission of closed strings
\footnote{ We note that both the annulus amplitude and the one point function  of a closed string vertex operator
on the disk are not suppressed by the diminishing string coupling since they are of order $g_s^0$.}. 
To obtain the precise form of the source, we examine the coupling of the brane to the dilaton fluctuations $\delta \Phi$ in the linearized approximation. This is given by
\begin{equation}
\int dtdx \sqrt{-g}e^{-2\Phi} \rho (x,t) (\delta \Phi),
\end{equation}
where 
\begin{eqnarray}
\rho(x,t) &=&  \frac{\delta(x-\bar{x}(t))}{l_s(c^2+\sinh^2t)^{1/2}}
\end{eqnarray}
and $\bar{x}(t)$ denotes the probe trajectory. Similar couplings are assumed for other closed string fields as well.
To get the amplitude relevant for particle
production, we need to Fourier transform the source with respect to the early, or
late time modes and evaluate it on shell, as described in section \ref{addsource}. 

We proceed to analyze the behavior of the Fourier transform 
of the source both in terms of the early and the late times modes. 
We have been able to compute the source in momentum space for the case $c=0$. This family of branes is special in that the branes remain localized in the $(x,t)$ coordinates and the time dependence results solely from the changing mass of the branes. 
Fourier transforming with respect to the early modes, we obtain
\begin{eqnarray}
{\tilde{\rho}}^{\ast} (j,p) &=& \nonumber \\
\int_{- \infty}^{+\infty} dx  \int_0^{\infty} dt \cosh t \sinh t
\delta (\sinh(x-x_0) \sinh t ) e^{ipx}
 & \nonumber \\
 (\sinh t)^{- i |p|} F(j+1-i |p|/2,-j-i|p|/2,-i|p|+1;-\sinh^2 t)
&=& \nonumber \\
\frac{1}{2} e^{i p x_0}
\Gamma (\frac{-is,is,-i|p|+1}{is-i|p|/2+1/2,-is -i|p|/2+1/2}), & & \label{ft1} 
\end{eqnarray}
while the Fourier transform with respect to the late modes yields
\begin{eqnarray}
{\tilde{\rho_l}}^{\ast} (j,p) &=& \nonumber \\
\int_{- \infty}^{+\infty} dx \int_0^{\infty} dt \cosh t \sinh t
\delta ( \sinh(x-x_0) \sinh t ) e^{ipx}
& & \nonumber \\
  (\sinh t)^{-2j-2} F(j+1+i|p|/2,j+1-i|p|/2,2j+2;-\sinh^{-2} t)
&=& \nonumber \\
\frac{1}{2} e^{i p x_0}
\Gamma 
(\frac{is,1+2is,1/2+i|p|/2,1/2-i|p|/2}{
is+i|p|/2+1/2,is -i|p|/2+1/2,1+is}). & & \label{ft2}
\end{eqnarray}
The phase factor $e^{ipx_0}$ in the amplitudes (\ref{ft1}) and (\ref{ft2}) can be simply understood as a consequence of translational invariance along the $x$-direction. The localized brane breaks this symmetry spontaneously.
For a discussion of the Fourier transform of the 
 generic source (and the comparison to the exact one-point function
for D1-branes in the euclidean cigar geometry), we refer to appendix \ref{genericFT}.

We can now turn to the computation of the number of on-shell particles produced
by the D0-brane. The generic result in equation
(\ref{partprod})
contains the bulk term, which we have discussed in detail in a previous section, and
the source term. The latter is relatively suppressed by an extra factor proportional to
the inverse volume of the spatial $x$-direction. This is because the source is localized in this
direction (and, in contrast to the bulk particle production, the source production does not occur
homogeneously in space). For the number of particles produced by the source, we obtain
\begin{eqnarray}
 \int {dp \over 2\pi} 
\frac{1}{2 |p|} |\tilde{\rho}(p) {{\alpha^{\ast}}} (p) +
\tilde{\rho}^{\ast} (-p) {\beta(-p)}^{\ast} |^2 
&=& \nonumber \\
\int {dp \over 2\pi}\left[\frac{\pi}{8 s^2} \frac{\cosh [ \pi(s+|p|/2) ] \cosh [ \pi(s-|p|/2) ] }{\sinh (2 \pi s)
  \cosh^2 (\pi |p|/2) }\right]. \label{sourcepartprod}
\end{eqnarray}
The energies and momenta satisfy the on-shell condition equation (\ref{onshell}).
The computation can either be done by using the Fourier transform with respect
to the early modes, and then using the formalism of section \ref{addsource},
or equivalently, by using directly the Fourier transform with respect to the
late modes (which automatically takes into account the cosmological evolution 
encoded in the Bogoljubov coefficients). 

The following comments are in order:
\begin{itemize}
\item In the limit of small $x$-momentum, the energy is essentially given by the effective cosmological mass, and the amplitude squared becomes ${|\rho_l|^2/ 4s} \sim {\coth ({m_{eff}/2}) \over m_{eff}^2}$. Thus for massive particles the production amplitude due to the brane source in this regime is of order one, while for massless particles we get an
 infrared divergence. The latter can be associated with the static long range fields produced by the source. The effective cosmological mass depends both on the momenta along the extra directions and the oscillator level of the emitted string state.
\item 
In the limit of large $x$-momentum and fixed effective cosmological mass, the on shell condition is given by $\omega = 2s \sim |p|$. In this limit we obtain
an extra exponential suppression factor $|\rho_l|^2/4s \sim \omega^{-2} e^{-  \pi \omega}$ in the amplitude of particle production. Notice that the exponent of the suppression factor is only half of that corresponding to bulk pair production in the large $x$-momentum regime. 
\item
When we embed in string theory, we need to take into account the enhancement
factor due to the exponentially large density of states at large energy. Now
in the case of the brane source, the closed string states that are produced
are left/right symmetric and their density is controlled by the open string
density \cite{Lambert:2003zr}. 
\end{itemize}

The considerations above indicate that the spectrum of closed strings emitted by the brane consists predominantely of small momentum highly massive states. If the brane is wrapped along extra flat directions, conservation of momentum along these directions forbids emission of strings with non-zero momentum along these extra directions. And for localized branes along flat extra directions, the amplitudes (for scalars) are modified by phase factors such as $e^{iky_0}$, with the dependence of the production amplitude squared on the momentum along these directions arising through the effective cosmological mass. As we saw, for higher effective mass there is a power law suppression factor. Since then there is an exponentially growing density of closed string states as a function of the oscillator level, the spectrum consists mostly of slowly moving, highly massive particles. Hence the emitted matter remains localized close to the brane. Presumably the mass gained by the brane, after backreaction is included, results in the production of such massive strings. For massive string states, it is important to consider finite $k$ corrections to the mini-superspace analysis we described. We comment on how to embed in string theory further below.     

As can be seen from the analysis in appendix B (see equation (\ref{c})), the
small $x$-momentum regime of the $c\ne 0$ branes is better in that an
exponential suppression factor appears for high cosmological mass (and hence
energy) given by $e^{-\pi \omega(1 -2q/\pi)}$, where the parameter $q$ is
defined by $\cos q =|c|$. For $c=0$, $q=\pi/2$, and this factor is one, while
for $ |c| <1$, $2q/\pi <1$, and an exponential suppression factor
survives. The reason for this phenomenon can be understood intuitively as
follows. When $c\ne 0$, these branes enter the cosmological region with
initial $x$-momentum, and the resulting radiation of strings with zero
$x$-momentum should now be less favored relatively to the emission of non-zero
momentum strings, always in comparison for the relative strengths in the $c=0$
case. 
%However, the suppression factor is not small enough to compete with the
%exponentially large Hagedorn
% density of closed string states in most string theory models one can construct.
For $|c|>1$, the parameter $q$ is imaginary and appears through an oscillatory
factor in the amplitude squared. For these cases the suppression factor seems
to be the biggest possible. We note that the trajectories of these branes,
when continued to the other
 regions of the cosmology, cross the timelike naked singularities.

\subsection*{Embedding in string theory}
Again, we can embed the model in string theory by adding a euclidean
black hole factor and extra flat space factors to the target space. It is
then straightforward to identify a region of phase space in which the energy grows like
the oscillator level (with all momenta fixed (and small)), where the particle production amplitude
is not suppressed significantly, and where the exponential degeneracy of states leads to
an exponentially divergent number of particles produced. As we pointed out
above, 
the exponential growth of the relevant states is governed by the open
string density. (One should compare this to the less
catastrophic behavior observed in \cite{Lambert:2003zr} for the case of unstable branes.)

When we perform the exact analysis beyond mini-superspace, we observe that an
extra factor (see appendix \ref{genericFT} and \cite{Ribault:2003ss})
\begin{eqnarray}
\Gamma(1 + \frac{i\omega}{k-2})
\end{eqnarray}
appears in the amplitude for one-closed string production, which leads
to an extra  suppression factor $e^{- \frac{ \pi \omega}{-k+2}}$ in 
the particle
production rate for $k$ negative. Sadly, this will generically be insufficient to lead
to finite particle production rates.
 There is a narrow window of opportunity, at very small $k$ (close to $2$),
i.e. a highly curved cosmology of the type discussed
in \cite{Hikida:2004mp}, in which one could swamp the Hagedorn 
density with this suppression factor. 
%It is interesting to note that no
%such opportunity appeared in the closed string sector.

Notice again the similarity of the resulting particle spectrum to the
 ``tachyon matter'' produced during the decay of unstable D-branes
 in flat space \cite{Sen:2002in}. It would be interesting
 to see if we can identify a tachyonic mode in the effective
 field theory on the brane (compare  e.g. to \cite{Yogendran:2004dm}) 
and see if we can map the problem
 to a tachyon condensation problem on the brane as in \cite{Kutasov:2004dj}.

\section{Conclusions}
\label{conclusions}
We have argued for an analytic continuation of the euclidean
$SL(2,R)/U(1)$ conformal field theory and its interpretation as the
future part of an interesting two-dimensional cosmology. We thus manage to
 incorporate all $\alpha'$ corrections to the relevant amplitudes discussed. 
We have identified and discussed a generic bulk backreaction problem
due to the Hagedorn degeneracy of string states.
The introduction of a non-perturbative source, such as a D0-brane, leads to 
additional particle production, which might be kept
under control in highly curved backgrounds. The formalism we developed
for the interplay between a cosmology and a time-dependent source should
be useful for other stringy cosmological models as well.

We have touched upon many directions of future research in the bulk of our paper.
Other regions of the Lorentzian cosmology (and D-branes propagating in these
regions) can be continued to other euclidean
conformal field theories, which in turn await an interpretation in the Lorentzian
realm (e.g. the D1-branes in the minimal
models). The patching together of different ``euclidean conformal
 field theory regions'' after suitable analytic continuation, and an understanding
of the matching of D-brane trajectories on the boundaries of these regions is an
interesting problem.
An easier first step might be found in the analysis of the boundary conditions induced
at the timelike singularities by particular euclidean conformal field theories. 

More generically, one can investigate the feasibility of formulating string field theory in
cosmological backgrounds, the analysis of the early vacuum,
and the relationship between open and closed string vacua 
induced by non-trivial consistency relationships between bulk and
boundary euclidean conformal field theories.

\section*{Acknowledgments}
We thank Ben Craps, Costas Kounnas, David Kutasov, Boris Pioline, 
Sylvain Ribault, Kazuhiro Sakai, Volker Schomerus and especially
Dan Israel for useful comments and lively discussions. We would moreover like
to thank Costas Kounnas for correcting a remark in an earlier version of this
paper. NT would like to thank
the department
 of theoretical physics at Ecole Normale Sup\'erieure for hospitality during which part of this work was completed.

\appendix
\section{Formalities}
\label{formal}
\subsection*{Formulas}
We introduce the following notation for a product
and quotient of Gamma-functions
\begin{eqnarray}
\Gamma \left( \frac{t_1,t_2, \dots, t_n}{b_1,b_2, \dots,b_m } \right)
&=& \frac{\Gamma(t_1)\Gamma(t_2) \dots \Gamma(t_n)}{\Gamma(b_1) \Gamma(b_2)
  \dots
\Gamma(b_m)}.
\end{eqnarray}
We use the transformation rule (i.e. rule for analytic continuation)
 for hypergeometric functions
\begin{eqnarray}
 (-z)^{-a} F(a,a+1-c,a+1-b;z^{-1}) &=& \nonumber \\
\Gamma(\frac{1-c,a+1-b}{1-b,a+1-c}) F(a,b,c;z) &  &
\nonumber \\
- \Gamma(\frac{c,1-c,a+1-b}{2-c,c-b,a}) e^{i \pi(c-1)}
z^{1-c} F(a+1-c,b+1-c,2-c;z), 
& & ,
\end{eqnarray}
and we make good use of the integral formulas \cite{GR}
\begin{eqnarray}
\int_0^{\infty} F(a,b,c;-z) z^{-t-1} dz &=&
\Gamma(\frac{a+t,b+t,c,-t}{a,b,c+t})
\end{eqnarray}
which
is valid when $Re(t)<0$, $Re(a+t)>0$, $Re(b+t)>0$ and $c$ should
not be a negative integer, and
\begin{eqnarray}
\int_0^{\infty} x^{c-1} (x+z)^{-\sigma} F(a,b,c;-x) dx
=
& & \nonumber \\
 \Gamma(\frac{c,a-c+\sigma,b-c+\sigma}{\sigma,a+b-c+\sigma}) 
F(a-c+\sigma,b-c+\sigma,a+b-c+\sigma;1-z) & & 
\end{eqnarray}
which holds when $Re(c) >0$, $Re(a-c+\sigma)>0$, $Re(b-c+\sigma)>0$ and
$|arg (z)| < \pi$.

\subsection*{Completeness}
We assume the following
completeness/orthogonality relations for the early modes (which should be
compared
to analogous relations for their euclidean counterparts \cite{Teschner:1997fv}):
\begin{eqnarray}
\rho(y,x) = \int \frac{dp}{2 \pi} \int_{-\frac{1}{2}+iR^+} \frac{dj}{2 \pi} 
                      {1 \over |N(j,p)|^2}\left[{\tilde{\rho}(j,p)} e^{ipx} f^j_{|p|} (y)
               + c.c.\right]
& & \nonumber \\
 \tilde{\rho}(j,p) = \int {dydx \over 2}  \rho(y,x) 
e^{-i px}   (f^j_{|p|} (y))^{\ast} & & \nonumber \\
\int {dydx \over 2}  ( e^{ipx} f^j_{|p|} (y) ) (e^{-ip'x}   (f^{j'}_{|p'|} (y))^{\ast})
= {{|N(j,p)|}^2} (2 \pi)^2 \delta(j-j') \delta(p-p'), & & 
\end{eqnarray}
where we recall that $y=\sinh^2 t$. The factor $|N(j,p)|$ is 
a normalization factor and the measure $\sqrt{-g}e^{-2\Phi}dt=dy/2$.

\section{The generic source in momentum space}
\label{genericFT}
In this appendix, we make some observations that lead to a guess for
an expression for the source in momentum space for a brane with a generic
trajectory. We will base our guess on the expression for the one-point function
of the D1-brane on the cigar geometry, which is the natural euclidean analog of our
D0-brane in the cosmology, as can  be seen by comparing their
trajectories. 

Before turning to this problem, we note an additional partial result.
We can determine the Fourier transform of the source at zero momentum ($p=0$) for 
a generic trajectory (parametrized by $c$) as follows
\begin{eqnarray}
\int_{- \infty}^{+\infty} dx \int_0^{\infty}dt \cosh t \sinh t
\delta (\sinh(x-x_0) \sinh t-c ) e^{ipx} 
& & \nonumber \\
(\sinh t)^{- i |p|} F(j+1-i |p|/2,-j-i|p|/2,-i|p|+1;-\sinh^2 t)
& & \nonumber \\
%&=& \nonumber 
%\end{eqnarray}
%reduces at $p=0$ to:
%\begin{eqnarray}
%\frac{1}{2} \int_0^{\infty} dz (z+|c|^2)^{-1/2} F(j+1,-j1;z)
%&=& \nonumber \\
 %\frac{1}{2} \Gamma(\frac{1,j+1/2,-j-1/2}{1/2,1/2})
%F(j+1/2,-j-1/2,1/2;1-|c|^2). & & 
%\end{eqnarray}
%We can then use the special form of the arguments of the hypergeometric
%function to rewrite this in terms of $\cos q = c$ as:
%\begin{eqnarray}
=_{|_{p=0}} \frac{1}{2 s} \frac{\cos (2j+1) q}{\sinh \pi s}, & & \label{c}
\end{eqnarray}
where we defined $\cos q = |c|$ for $|c|<1$ and $q \in {[} 0 , \pi/2 {]}$.

With this extra trump in our hand, we turn to the full problem.
We note that the semi-classical
one-point function for the D1-brane in
the cigar \cite{Ribault:2003ss}
can be rewritten, after multiplying the
one-point function by the
inverse
of the normalization factor for the euclidean eigenfunctions, as
\begin{eqnarray}
< \chi_{closed} > &=&  e^{i n \theta_0} \frac{\sin \pi(\frac{1}{2}-is+|n|/2)}{2 s \sinh 2 \pi s}
\Gamma(\frac{|n|+1}{1/2+is+|n|/2,1/2-is+|n|/2})\nonumber \\
& &  \left( e^{-r(2j+1)}+e^{i \pi n} e^{r
  (2j+1)}  \right),
\end{eqnarray}
which at $r=0$ can be rewritten, for $n$ even, as
\begin{eqnarray}
< \chi_{closed} > &=&  e^{i n (\theta_0+ \frac{\pi}{2})}  \frac{1}{2s \sinh \pi s}
\Gamma(\frac{|n|+1}{\frac{1}{2}+is+|n|/2,\frac{1}{2}-is+|n|/2}).
\end{eqnarray}
We expect the analytic continuation to map $|n|$ to $-i|p|$ (since
this is the prescription that maps normalizable euclidean modes to early
positive frequency modes), and indeed, we find a match with our results for
the Fourier transform at $c=0$ (up to an overall constant).

To address the generic case, it is important to notice that 
the D1-brane on the cigar intersects most circular cross sections of the
cigar in precisely two points, while the D0-brane in the cosmology only
intersects
a spatial slice at one point. Naively, analytically continuing the cigar
source
to the Lorentzian theory would lead to a doubling of D-branes.

To obtain the correct one-point function for a single source, we propose to 
perform the following rewriting of the one-point function
\begin{eqnarray}
<\chi_{closed}> &=& e^{+ i \pi n/2} e^{i n \theta_0}
\frac{1}{2 s \sinh 2 \pi s} \Gamma(\frac{|n|+1}{1/2+is+|n|/2,1/2-is+|n|/2})
\nonumber \\
& & 
\left( 
\cosh s(\pi +2ri) e^{i n \pi} + \cosh s(\pi -2ri) 
\right).
\end{eqnarray}
We wish to identify the two terms with two possible sources in the
Lorentzian theory, at $c$ and $x_0=i(\theta_0+\pi)$ and at $-c$ and 
$x_0=i \theta_0$ 
respectively.
% Notice
% that the two terms are exchanged by the parity operation $(\theta_0, r, n)
% \to (-\theta_0, -r, -n)$ which is expected to relate the amplitudes for the
% two sources at $|c|$ and $-|c|$ naturally. 
The results now match onto the
results we obtained for generic $|c|=\cos q$ at $p=0$, taking into account that $c=\sin r'$ and $q=\pi/2-|r'|$, for
$r'=ir \in {[} -\pi/2, \pi/2 {]}$, and taking into account just the corresponding single source term.

Although we haven't been able to do the integral exactly, we thus propose
an expression for the FT of the generic case (depending on the sign of $r'$):
\begin{eqnarray}
e^{i p x_0} \frac{1}{4 s \sinh \pi s \cosh \pi s}  
\Gamma(\frac{-i|p|+1}{1/2+is-i|p|/2,1/2-is-i|p|/2})
\cosh \pi s(1 \mp r' \frac{2}{\pi}) .
\end{eqnarray}
We estimate that the generic integral that should lead to this result
 can presumably be performed by delving
more
deeply into the representation theory of $SL(2,R)$.
We observe that the large energy behavior of the one-point function
crucially depends on the value of the parameter $r'$. 
%In particular,
%the last factor in the one-point function is exponentially growing
%or damped depending on whether $|r'|$ is smaller or greater than
%$\pi/2$. We observe that
%for a value of $|r'|$ greater than $\pi/2$, the D0-brane
%trajectory crossed the (naive) space-time singularities in the cosmology
%in the past.

This guess for the full semi-classical one-point function
can now be extended to the exact one-point function for
the D0-brane in the cosmology, by repeating the above
manipulations on the known \cite{Ribault:2003ss}
exact result for the one-point function of the D1-brane in 
the euclidean cigar . This leads to the extra factor in the exact result 
which
we quoted in the bulk of our letter.

%\section{A few quick computations}
%Consider the analytically continued normalizable euclidean modes:
%\begin{eqnarray}
%u^{e}_p &=& \frac{1}{\sqrt{2|p|}} e^{ipx} 
%y^{-i|p|/2} F(j+1-i|p|/2,-j-i|p|/2,-i|p|+1;-y). 
%\end{eqnarray}
%They are proportional at large $t,y$ to:
%\begin{eqnarray}
%u_1 ~ u_p^e & ~ & e^{ipx} y^{-j-1} \Gamma(-i|p|+1,-2j-1 / -j-i|p|/2,-j-i|p|/2)
%\nonumber \\
%& & 
%+ e^{ipx} y^{j} \Gamma(-i |p|+1,2j+1 / j+1-i|p|/2,j+1-i|p|/2).
%\end{eqnarray}
%We normalize the first factor such as to eliminate the $\Gamma$-factors,
%and we find (see HT) the semi-classical reflection coefficient:
%\begin{eqnarray}
%c &=& \Gamma(2j+1,-j-i|p|/2,-j-i|p|/2 / -2j-1, j+1-i|p|/2,j+1-i|p|/2 ).
%\end{eqnarray}
%Comparing to the exact result in RS, we see that we miss the prefactor 
%$\nu$, and the last factor-fraction (in RS (2.13)). The last factor (which is
%a phase) is  a stringy
%effect (HT), and the $\nu$ factor, after analytic continuation, and
%for negative level is (also) a phase (since $\nu$ is then positive).
%We can check that the following is true (compare to ST page 372):
%\begin{eqnarray}
%|c|^2 &=& \frac{|\beta|^2}{|\alpha|^2}.
%\end{eqnarray}

%

\begin{thebibliography}{99}

%\cite{Gutperle:2003xf}\cite{Strominger:2003fn}\cite{Schomerus:2003vv}\cite{Fredenhagen:2003ut}
\bibitem{Gutperle:2003xf}
M.~Gutperle and A.~Strominger,
%``Timelike boundary Liouville theory,''
Phys.\ Rev.\ D {\bf 67} (2003) 126002
[arXiv:hep-th/0301038].
%%CITATION = HEP-TH 0301038;%%
%\cite{Strominger:2003fn}
\bibitem{Strominger:2003fn}
A.~Strominger and T.~Takayanagi,
%``Correlators in timelike bulk Liouville theory,''
Adv.\ Theor.\ Math.\ Phys.\  {\bf 7}, 369 (2003)
[arXiv:hep-th/0303221].
%%CITATION = HEP-TH 0303221;%%
%\cite{Schomerus:2003vv}\cite{Fredenhagen:2003ut}
\bibitem{Schomerus:2003vv}
V.~Schomerus,
%``Rolling tachyons from Liouville theory,''
JHEP {\bf 0311} (2003) 043
[arXiv:hep-th/0306026].
%%CITATION = HEP-TH 0306026;%%
%\cite{Fredenhagen:2003ut}
\bibitem{Fredenhagen:2003ut}
S.~Fredenhagen and V.~Schomerus,
%``On minisuperspace models of S-branes,''
JHEP {\bf 0312} (2003) 003
[arXiv:hep-th/0308205].
%%CITATION = HEP-TH 0308205;%%

%\cite{Kutasov:2004dj}
\bibitem{Kutasov:2004dj}
D.~Kutasov,
%``D-brane dynamics near NS5-branes,''
arXiv:hep-th/0405058.
%%CITATION = HEP-TH 0405058;%%

%\cite{Nakayama:2004yx}
\bibitem{Nakayama:2004yx}
Y.~Nakayama, Y.~Sugawara and H.~Takayanagi,
%``Boundary states for the rolling D-branes in NS5 background,''
JHEP {\bf 0407}, 020 (2004)
[arXiv:hep-th/0406173].
%%CITATION = HEP-TH 0406173;%%

%\cite{Kounnas:1992wc}\cite{Cornalba:2002nv}\cite{Craps:2002ii}
\bibitem{Kounnas:1992wc}
C.~Kounnas and D.~Lust,
%``Cosmological string backgrounds from gauged WZW models,''
Phys.\ Lett.\ B {\bf 289}, 56 (1992)
[arXiv:hep-th/9205046].
%%CITATION = HEP-TH 9205046;%%
%\cite{Cornalba:2002nv}\cite{Craps:2002ii}
\bibitem{Cornalba:2002nv}
L.~Cornalba, M.~S.~Costa and C.~Kounnas,
%``A resolution of the cosmological singularity with orientifolds,''
Nucl.\ Phys.\ B {\bf 637} (2002) 378
[arXiv:hep-th/0204261].
%%CITATION = HEP-TH 0204261;%%
%\cite{Craps:2002ii}
\bibitem{Craps:2002ii}
B.~Craps, D.~Kutasov and G.~Rajesh,
%``String propagation in the presence of cosmological singularities,''
JHEP {\bf 0206}, 053 (2002)
[arXiv:hep-th/0205101].
%%CITATION = HEP-TH 0205101;%%
%\cite{Ribault:2003ss}\cite{Israel:2004jt}\cite{Fotopoulos:2004ut}
\bibitem{Ribault:2003ss}
S.~Ribault and V.~Schomerus,
%``Branes in the 2-D black hole,''
JHEP {\bf 0402}, 019 (2004)
[arXiv:hep-th/0310024].
%%CITATION = HEP-TH 0310024;%%
%\cite{Israel:2004jt}\cite{Fotopoulos:2004ut}
\bibitem{Israel:2004jt}
D.~Israel, A.~Pakman and J.~Troost,
%``D-branes in N = 2 Liouville theory and its mirror,''
arXiv:hep-th/0405259.
%%CITATION = HEP-TH 0405259;%%
%\cite{Fotopoulos:2004ut}
\bibitem{Fotopoulos:2004ut}
A.~Fotopoulos, V.~Niarchos and N.~Prezas,
%``D-branes and extended characters in SL(2,R)/U(1),''
arXiv:hep-th/0406017.
%%CITATION = HEP-TH 0406017;%%

\bibitem{BD}
N. Birrell and P. Davies, Quantum fields in curved space, Cambridge University
Press 1984.
%\cite{Elitzur:2002rt}
\bibitem{Elitzur:2002rt}
S.~Elitzur, A.~Giveon, D.~Kutasov and E.~Rabinovici,
%``From big bang to big crunch and beyond,''
JHEP {\bf 0206} (2002) 017
[arXiv:hep-th/0204189].
%%CITATION = HEP-TH 0204189;%%

\bibitem{BZ}
B. Zwiebach, A first course in string theory, Cambridge University Press 2004.

%\cite{Hikida:2004mp}
\bibitem{Hikida:2004mp}
Y.~Hikida and T.~Takayanagi,
%``On solvable time-dependent model and rolling closed string tachyon,''
arXiv:hep-th/0408124.
%%CITATION = HEP-TH 0408124;%%

%\cite{Karczmarek:2003xm}
\bibitem{Karczmarek:2003xm}
J.~L.~Karczmarek, H.~Liu, J.~Maldacena and A.~Strominger,
%``UV finite brane decay,''
JHEP {\bf 0311}, 042 (2003)
[arXiv:hep-th/0306132].
%%CITATION = HEP-TH 0306132;%%

\bibitem{PS}
M. Peskin and D. Schr\"oder, An introduction to quantum field theory,
Addison-Wesley 1997.
H 9711200;%%

%%CITATIO%\cite{Yogendran:2004dm}
\bibitem{Yogendran:2004dm}
K.~P.~Yogendran,
%``D-branes in 2D Lorentzian black hole,''
arXiv:hep-th/0408114.
%%CITATION = HEP-TH 0408114;%%N = HEP-T

%\cite{Sen:2002in}
\bibitem{Sen:2002in}
A.~Sen,
%``Tachyon matter,''
JHEP {\bf 0207}, 065 (2002)
[arXiv:hep-th/0203265].
%%CITATION = HEP-TH 0203265;%%

%\cite{Lambert:2003zr}
\bibitem{Lambert:2003zr}
N.~Lambert, H.~Liu and J.~Maldacena,
%``Closed strings from decaying D-branes,''
arXiv:hep-th/0303139.
%%CITATION = HEP-TH 0303139;%%

%\cite{Sen:2003xs}
\bibitem{Sen:2003xs}
A.~Sen,
%``Open-closed duality at tree level,''
Phys.\ Rev.\ Lett.\  {\bf 91}, 181601 (2003)
[arXiv:hep-th/0306137].
%%CITATION = HEP-TH 0306137;%%






%\cite{Gubser:2003vk}
\bibitem{Gubser:2003vk}
S.~S.~Gubser,
%``String production at the level of effective field theory,''
Phys.\ Rev.\ D {\bf 69}, 123507 (2004)
[arXiv:hep-th/0305099].
%%CITATION = HEP-TH 0305099;%%


%\cite{Friess:2004zk}
\bibitem{Friess:2004zk}
J.~J.~Friess, S.~S.~Gubser and I.~Mitra,
%``String creation in cosmologies with a varying dilaton,''
Nucl.\ Phys.\ B {\bf 689}, 243 (2004)
[arXiv:hep-th/0402156].
%%CITATION = HEP-TH 0402156;%%




%\cite{Teschner:1997fv}
\bibitem{Teschner:1997fv}
J.~Teschner,
%``The mini-superspace limit of the SL(2,C)/SU(2) WZNW model,''
Nucl.\ Phys.\ B {\bf 546}, 369 (1999)
[arXiv:hep-th/9712258].
%%CITATION = HEP-TH 9712258;%%

\bibitem{GR}
I. Gradshteyn and I. Ryzhik, Table of integrals, series and products, Academic
Press 1965.
\end{thebibliography}
\end{document}